\documentclass[aps,prl,twocolumn,amsmath,amssymb,amsfonts,nofootinbib,long,floatfix]{revtex4}
\usepackage{epsfig,latexsym,bm,epstopdf}

\usepackage{color}

\makeatletter
\newcommand{\vast}{\bBigg@{4}}
\newcommand{\Vast}{\bBigg@{5}}
\makeatother

\DeclareMathAlphabet{\mathpzc}{OT1}{pzc}{m}{it}

\allowdisplaybreaks[4]
\usepackage{amsmath}
\usepackage{amssymb}

\begin{document}

\title{Detection of extremely low frequency gravitational wave using gravitational lens: The general case}

\author{Wenshuai Liu$^{1}$}
\email{674602871@qq.com}
\affiliation{$^1$School of Physics, Huazhong University of Science and Technology, Wuhan 430074, China}

\date{\today}


\begin{abstract}
The effect of gravitational wave of extremely low frequency on time delays between different locations on the Einstein ring in a lens system with an aligned source-deflector-observer configuration is investigated. The observer will observe an Einstein ring from the lens system aligned in a highly symmetric configuration. Time delays between different locations on the Einstein ring cannot emerge without gravitational wave of cosmological wavelength propagating through the lens system. Otherwise, different locations on the Einstein ring will show time delays. Our previous studies showed that time delays from different locations on the Einstein ring in the presence of gravitational wave of cosmological wavelength present a special relationship. But this result is limited to a specific aligned source-deflector-observer configuration where the source and the observer are equidistant from the deflector and the gravitational wave has special direction of propagation and polarization. In order to investigate the property in the general case, we extend to the general condition that the source and the observer aren't equidistant from the deflector and the direction and polarization of the gravitational wave is arbitrary in the present work. Results in this work show that time delays between different locations on the Einstein ring with the general condition present the same relationship as that in our previous studies.
\end{abstract}




\maketitle


\section{Introduction}

Inflation, according to which the initial perturbations giving rise to the structure in the Universe originated from quantum fluctuations during a period of accelerating expansion in the very early Universe, is the leading theory which leads to a flat, homogeneous and isotropic Universe \cite{9,10,11,12}. In addition, inflation predicts the existence of primordial gravitational waves (PGWs) with a nearly scale-invariant spectrum \cite{1,2,3,4,5,6,7}. The observation of PGWs will not only verify the inflationary theory but also provide the information about the energy scale of inflation. PGWs with cosmological wavelength would induce the B-mode of polarization of the cosmic microwave background (CMB) via Thomson scattering and B-mode is widely considered as the method of detecting PGWs with extremely low frequency in the range of $10^{-18}$Hz-$10^{-16}$Hz \cite{13,14}.

In order to provide a positive independent confirmation of PGW besides its signature on the polarization of CMB, it is of great significance to investigate an alternative observational feature originating from PGW. A characteristic pattern of the source's apparent proper motion emerges as a result of propagation of light traveling through gravitational waves, showing oscillations in the source's apparent position with the same period as that of gravitational wave \cite{1803}. Detecting gravitational waves emitted by binaries through microlensing has been investigated by \cite{1804,1805}. Some other indirect ways to detect PGW include an upper limit on the stochastic gravitational wave background using galaxy-galaxy n-point correlation functions \cite{17}, the proper motion of distant astrophysical objects induced by gravitational wave \cite{18,1801,1802} and gravitational lensing of the CMB \cite{19}.

Recently, Liu \cite{1901} proposed a new method to detect PGW with extremely low frequency, showing that gravitational lens could serve as a potential PGW detector. In Liu's work \cite{1901}, time delays with a special relationship shown in Eq. (26) in Liu \cite{1901} between different locations on the Einstein ring will emerge if extremely low frequency PGW propagates through the lens system where the source and the observer are equidistant from and aligned with the deflector in a highly symmetric configuration with the condition that $\omega L \eta \ll 1$ and $h \ll \eta$ where $L$ is the distance from the source (or the observer) to the deflector and $\eta$ is the Einstein radius, and $\omega$, $h$ represent the frequency and the dimensionless amplitude of the gravitational wave, respectively. Such detection of extremely low frequency PGW using gravitational lens is proposed based on \cite{20,15,21} where Allen \cite{20,15} suggested that gravitational lenses could act as detectors to probe extremely low frequency PGW with time delays between different images of a quasar and Frieman \cite{21} later on demonstrated that the way proposed by Allen \cite{20,15} could not work due to the fact that time delays induced by PGW cannot be observationally distinguishable from the intrinsic time delays originating from the geometry of the gravitational lens. Liu \cite{1901} proved that gravitational lens with an aligned source-deflector-observer configuration could detect extremely low frequency PGW when the whole Einstein ring is taken into account.

In this work, in order to investigate whether the result in \cite{1901} holds in the general condition, we expand the condition adopted by Liu \cite{1901} that the source and the observer are equidistant from the deflector and the gravitational wave has a specific polarization. Here, we assume the distance from the observe to the deflector is not equal to that from the source to the deflector and that the polarization of the gravitational wave is arbitrary. The results show that time delays between different locations on the Einstein ring have the same relationship as that shown in Eq. (26) in Liu \cite{1901}. This work presents that time delays with this special relationship are independent of the ratio of the distance from the observe to the deflector to the distance from the source to the deflector, the direction of propagation and polarization of PGWs, and the linear superposition of PGWs, meaning that extremely low frequency PGWs are confirmed if time delays from different locations on the Einstein ring show such relationship.

\section{Time delay due to primordial gravitational waves}

\begin{figure}[t!]
\begin{center}
\includegraphics[clip,width=0.5\textwidth]{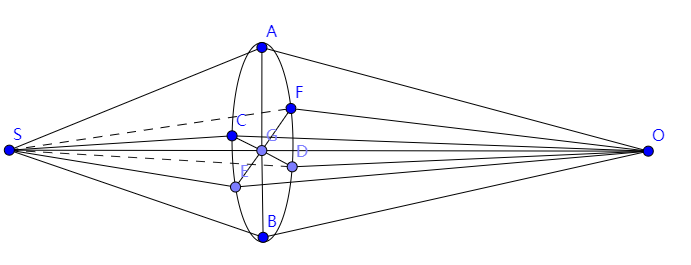}
\caption{ $S$, $G$ and $O$ represent the source, the deflector and the observer, respectively. $S$, $G$ and $O$ are on the z axis with coordinates $S$ $(x=0,y=0,z=-D_{LS})$, $G$ $(x=0,y=0,z=0)$ and $O$ $(x=0,y=0,z=D_L)$. $A$, $B$, $C$ and $D$ are on the Einstein ring.}
\end{center}
\end{figure}

In order to present the aim of this work clearly, we show the same lens geometry as that in Frieman \cite{21} with an aligned source-deflector-observer configuration shown in Figure 1. A point (or thin axially symmetric) gravitational deflector with mass $M$ is at the origin of coordinates in the lens configuration.We set the distance from the observer to the deflector to be $D_L$ and that from the deflector to the source to be $D_{LS}$. Thus, the source (quasar) and the observer are located at $(x=0,y=0,z=-D_{LS})$ and $(x=0,y=0,z=D_L)$, respectively. With such gravitational lens geometry, the observer would see an Einstein ring image of the source with an angular radius of $\eta=\sqrt{\frac{4GMD_{LS}}{D_L(D_L+D_{LS})}}$ (where we set the speed of light $c=1$) without time delays between different locations on the Einstein ring if there is no gravitational wave propagating through the lens system. We set $D_L=L$ and assume $R=\frac{D_{LS}}{D_L}$ is arbitrary. We denote the gravitational potential of the deflector to be $U$ and consider a gravitational wave propagating through the lens system where we assume the polarization and direction of propagation of the gravitational wave is arbitrary and that direction of propagation lies in the $x-z$ plane. Finally, the metric induced by the gravitational wave has the following form
\begin{eqnarray}
h_{ij}=&&
\left[\begin{array}{cccc}
    H_{11} &    H_{12}    & H_{13} \\
    H_{21} &    H_{22}   & H_{23}\\
    H_{31} & H_{32} & H_{33}
\end{array}\right]
\nonumber \\
&& \times  \cos(\omega t-\mathbf{k} \cdot \mathbf{x})
\end{eqnarray}
where $\mathbf{k}=\omega(\sin\phi,0,\cos\phi)$ is the propagation vector, $\omega$ is the frequency, $h_+$ and $h_\times$ are the amplitude of the two polarizations of the gravitational wave, respectively.

Same as Liu \cite{1901}, we get the total metric given by
\begin{equation}
ds^2=(1+2U)dt^2-(1-2U)(dx^2+dy^2+dz^2)+h_{ij}dx^idx^j \label{1}
\end{equation}

and the time of travel of light
\begin{equation}
T \approx \int_{-D_{LS}}^{D_L}dz\left[1+\frac{1}{2}\left(\frac{dx}{dz}\right)^2+\frac{1}{2}\left(\frac{dy}{dz}\right)^2+\frac{1}{2}h_{ij}\frac{dx^i}{dz}\frac{dx^j}{dz}-2U\right] \label{2}
\end{equation}

The first three terms in the bracket of Eq. (\ref{2}) and the fifth arise from the lens geometry and the gravitational potential of the deflector, respectively. The contribution from gravitational wave to the time of travel is due to the fourth term. We replace $t$ in Eq. (\ref{2}) by $t=t_e+(z+D_{LS})$ for the level of approximation where $t_e$ is the time the photons were emitted at $(x=0,y=0,z=-D_{LS})$.

With the same method in Liu \cite{1901}, we approximate photon trajectories by straight segments shown in Figure 1. Then the photon trajectories are
\begin{eqnarray}
x_{1,2}=\pm\frac{\eta\cos{\theta}}{R}(z+RL), z<0 \label{3}\\
x_{1,2}=\mp\eta\cos{\theta}(z-L), z>0 \label{4}\\
y_{1,2}=\pm\frac{\eta\sin{\theta}}{R}(z+RL), z<0 \label{5}\\
y_{1,2}=\mp\eta\sin{\theta}(z-L), z>0 \label{6}
\end{eqnarray}
where the subscripts ${1,2}$ represent the trajectories of photons traveling through different locations on the Einstein ring, $(A,B)$, $(C,D)$, $(E,F)$.

As we can see from Eq. (\ref{2}) with the symmetry of the light paths, the only contribution to time delay for two locations which are symmetrical about the gravitational lens on the Einstein ring is from the fourth term in Eq. (\ref{2}). Integrating Eq. (\ref{2}) with Eqs. (\ref{3})-(\ref{6}) directly is complex and difficult. To make this question simple, we investigate the fourth term of the integrand for the two photon trajectories which are symmetrical about the lens axis.

For $h_{33}$ in the case of $z<0$, we get
\begin{eqnarray}
\Delta(\frac{1}{2}h_{33}\frac{dz}{dz}\frac{dz}{dz})=&&H_{33}\sin{[\omega(t_e+z+RL-\cos{\phi}z)]}
\nonumber \\
&&\times \sin{[\omega\sin{\phi}\cos{\theta}\frac{\eta}{R}(z+RL)]} \label{7}
\end{eqnarray}
where $\Delta(\frac{1}{2}h_{33}\frac{dz}{dz}\frac{dz}{dz})=\frac{1}{2}h_{33}\frac{dz}{dz}\frac{dz}{dz}|_1-\frac{1}{2}h_{33}\frac{dz}{dz}\frac{dz}{dz}|_2$ in which the subscripts ${1,2}$ represent the trajectories of photons shown in Eqs. (\ref{3})-(\ref{6}).

With the condition $\omega L \eta \ll 1$, Eq. (\ref{7}) can be simplified as
\begin{eqnarray}
\Delta(\frac{1}{2}h_{33}\frac{dz}{dz}\frac{dz}{dz})=&&H_{33}\sin{[\omega(t_e+z+RL-\cos{\phi}z)]}
\nonumber \\
&&\times [\omega\sin{\phi}\cos{\theta}\frac{\eta}{R}(z+RL)] \label{8}
\end{eqnarray}

In a similar way, we get the follows due to $h_{23}/h_{32}$, $h_{13}/h_{31}$
\begin{eqnarray}
\Delta(\frac{1}{2}h_{23}\frac{dy}{dz}\frac{dz}{dz})=&&H_{23}\frac{\eta\sin{\theta}}{R}\cos{[\omega(t_e+z+RL-\cos{\phi}z)]}
\nonumber \\
&&\times \cos{[\omega\sin{\phi}\cos{\theta}\frac{\eta}{R}(z+RL)]} \label{9}
\end{eqnarray}

\begin{eqnarray}
\Delta(\frac{1}{2}h_{13}\frac{dx}{dz}\frac{dz}{dz})=&&H_{13}\frac{\eta\cos{\theta}}{R}\cos{[\omega(t_e+z+RL-\cos{\phi}z)]}
\nonumber \\
&&\times \cos{[\omega\sin{\phi}\cos{\theta}\frac{\eta}{R}(z+RL)]} \label{10}
\end{eqnarray}

For $h_{11}$, $h_{12}$, $h_{21}$ and $h_{22}$, $\frac{1}{2}h_{ij}\frac{dx^i}{dz}\frac{dx^j}{dz}$ is negligible compared with that with $h_{33}$, $h_{13}$, $h_{31}$, $h_{23}$ and $h_{32}$.

Then, combined with Eqs. (\ref{8}), (\ref{9}) and (\ref{10}), it shows that
\begin{equation}
C_0=C_1\cos{\theta}+C_2\sin{\theta}\label{11}
\end{equation}
where
\begin{equation}
C_0=\Delta(\frac{1}{2}h_{ij}\frac{dx^i}{dz}\frac{dx^j}{dz})\label{12}
\end{equation}

\begin{eqnarray}
C_1=&&H_{33}\sin{[\omega(t_e+z+RL-\cos{\phi}z)]}
\nonumber \\
&&\times [\omega\sin{\phi}\frac{\eta}{R}(z+RL)]
\nonumber \\
&&+H_{13}\frac{\eta}{R}\cos{[\omega(t_e+z+RL-\cos{\phi}z)]}
\nonumber \\
&&\times \cos{[\omega\sin{\phi}\cos{\theta}\frac{\eta}{R}(z+RL)]}
\nonumber \\
&&+H_{31}\frac{\eta}{R}\cos{[\omega(t_e+z+RL-\cos{\phi}z)]}
\nonumber \\
&&\times \cos{[\omega\sin{\phi}\cos{\theta}\frac{\eta}{R}(z+RL)]}\label{13}
\end{eqnarray}
and
\begin{eqnarray}
C_2=&&H_{23}\frac{\eta}{R}\cos{[\omega(t_e+z+RL-\cos{\phi}z)]}
\nonumber \\
&&\times \cos{[\omega\sin{\phi}\cos{\theta}\frac{\eta}{R}(z+RL)]}
\nonumber \\
&&+H_{32}\frac{\eta}{R}\cos{[\omega(t_e+z+RL-\cos{\phi}z)]}
\nonumber \\
&&\times \cos{[\omega\sin{\phi}\cos{\theta}\frac{\eta}{R}(z+RL)]}\label{14}
\end{eqnarray}

Let's assume that $\theta$ is the angle between the line AGB and the projection of the direction of propagation of gravitational wave on the plane containing the Einstein ring, $\angle AGC = a$ is the angle between the line AGB and the line CGD and $\angle CGE = b$ is the angle between the line CGD and the line EGF, we obtain the following relationship between $C_{0(AB)}$, $C_{0(CD)}$ and $C_{0(EF)}$ based on Eq. (\ref{11})

\begin{eqnarray}
C_{0(CD)}=&&C_{0(AB)}\left[\cos a-\frac{\sin a \cos(a+b)}{\sin(a+b)}\right]
\nonumber \\
&&+C_{0(EF)}\frac{\sin a }{\sin(a+b)} \label{15}
\end{eqnarray}

Similarly, we could get the same equation as Eq. (\ref{15}) for $z>0$. Based on Eq. (\ref{15}), we obtain the following relationship

\begin{equation}
\Delta T_{CD}=\Delta T_{AB}\left[\cos a-\frac{\sin a \cos(a+b)}{\sin(a+b)}\right] +\Delta T_{EF}\frac{\sin a }{\sin(a+b)} \label{16}
\end{equation}
where $\Delta T_{\rm MN}=T_{\rm SMO}-T_{\rm SNO}$, $T_{\rm SMO}$ is the time travel of the light traveling from the source $S$ through the point $M$ on the Einstein ring to the observer $O$, $M/N$ represents $A$, $B$, $C$ and $D$ on the Einstein ring.

We investigate time delays induced by an extremely low frequency PGW with arbitrary direction of propagation and polarization throughout above. In the case of an aligned source-deflector-observer configuration with many extremely low frequency PGWs traveling through the lens system, Eq. (\ref{16}) holds according to the linear superposition of PGWs with extremely low frequency. It shows that Eq. (\ref{16}) presents the same result as that in Liu \cite{1901}, meaning that time delays from different locations on the Einstein ring with the relationship represented by Eq. (\ref{16}) are independent of the direction of propagation and polarization of PGWs, the linear superposition of PGWs, and the ratio of $D_{LS}$ to $D_L$, and that PGWs with extremely low frequency are detected if time delays (time delays approach hours according to \cite{1901}) from different locations on the Einstein ring show the relationship in Eq. (\ref{16}).

\section{Conclusion and discussion}
We investigate time delays located on the Einstein ring in an aligned lens configuration with the general condition, which extends the case studied in Liu \cite{1901} that the source and the observer are equidistant from the deflector and the PGW has a specific polarization. In this work, we adopt a general lens geometry where the ratio of the distance from the observe to the deflector to the distance from the source to the deflector is arbitrary. Based on this condition, the results demonstrate that time delays located on the Einstein ring with the relationship shown in Eq. (\ref{16}) are independent of the ratio of the distance from the observe to the deflector to the distance from the source to the deflector, the direction of propagation and polarization of PGWs and the linear superposition of PGWs, meaning that gravitational lens with the aligned source-deflector-observer configuration could serve as a potential PGW detector.

In order to use an Einstein ring to detect gravitational wave of cosmological wavelength, the observer must be able to distinguish gravitational wave induced time delays from the ones produced by large scale structure. Results from \cite{25} showed that an aligned lens in the presence of large scale structure fluctuations is equivalent to that of a similar lens with a nonaligned configuration and no perturbations from large scale structure. For different locations on the Einstein ring, the large scale structure cannot induced the special time delays with the relationship shown in Eq. (\ref{16}), meaning that relationship represented by Eq. (\ref{16}) is a unique feature resulting from an aligned source-detector-observer configuration with extremely low frequency PGWs and that extremely low frequency PGWs are detected if time delays from different locations on the Einstein ring present the relationship shown in Eq. (\ref{16}). Furthermore, once time delays from an Einstein ring with the relationship in Eq. (\ref{16}) are observed, one cannot tell how many superpositioned gravitational waves propagating through the lens system and what the direction of propagation and polarization of extremely low frequency PGWs in detail, but one could confirm that PGWs are detected.

Due to the low possibility of having gravitational lens with an aligned source-detector-observer configuration, only a handful of Einstein ring systems with complete or nearly complete ring morphology have been discovered up to now \cite{26,27,28}. It is expected that many gravitational lenses showing nearly perfect Einstein rings will be discovered with the advent of ultra-wide
cameras and large area surveys in future \cite{29,30}. The inner accretion disk around the central supermassive black hole in a quasar usually exhibits variable accretion which leads to variability in luminosity, providing an ideal source in the lens system. With the high time resolution in astronomical observation \cite{31,32}, it is possible to use Einstein rings to detect extremely low frequency PGWs produced by inflation.


\begin{thebibliography}{99}

\bibitem{9}L. P. Grishchuk, JETP Lett. 23, 293 (1976).

\bibitem{10}L. P. Grishchuk, Sov. Phys. Usp. 20, 319 (1977).

\bibitem{11}A. A. Starobinsky, Phys. Lett. B 91, 99 (1980).


\bibitem{12}A. D. Linde, Phys. Lett. B 108, 389 (1982).

\bibitem{1}L. F. Abbott \& M. B. Wise, Nucl. Phys. B244, 541 (1984).
\bibitem{2}A. Starobinskii, Sov. Astron. Lett. 11, 133 (1985).
\bibitem{3}V. A. Rubakov, M.V. Sazhin, \& A.V. Veryaskin, Phys. Lett. 115B, 189 (1982).

\bibitem{4}R. Fabbri \& M. D. Pollock, Phys. Lett. 125B, 445 (1983).
\bibitem{5}A. A. Starobinsky, JETP Lett. 30, 682 (1979).
\bibitem{6}V. Sahni, Phys. Rev. D 42, 453 (1990).
\bibitem{7}B. Allen, Phys. Rev. D 37, 2078 (1988).






\bibitem{13}M. Kamionkowski, A. Kosowsky, \& A. Stebbins, Phys. Rev. Lett. 78, 2058 (1997).

\bibitem{14}U. Seljak, \& M. Zaldarriaga, Phys. Rev. Lett. 78, 2054 (1997).


\bibitem{1803}T. Pyne, C. R. Gwinn, M. Birkinshaw, T. M. Eubanks, \&  D. N. Matsakis, Astrophys. J., 465, 566 (1996).




\bibitem{1804}S. L. Larson, \& R. Schild, ArXiv e-prints, arXiv:0007142 (2000).

\bibitem{1805}R. Ragazzoni, G. Valente, \& E. Marchetti, Mon. Not. R. Astron. Soc. 345, 100 (2003).


\bibitem{17}E. V. Linder, Astrophys. J., 328, 77 (1988).

\bibitem{18}R. Bar-Kana, Phys. Rev. D, 54, 7138 (1996).

\bibitem{1801}L. G. Book, \& {\'E}. {\'E}. Flanagan, Phys. Rev. D 83, 024024 (2011).

\bibitem{1802}S. Bharadwaj, \& T. Guha Sarkar, Phys. Rev. D, 79, 124003 (2009).





\bibitem{19}L. G. Book, M. Kamionkowski, \& T. Souradeep, Phys. Rev. D, 85, 023010 (2012).

\bibitem{1901}W. Liu, Phys. Rev. D, 103, 103012 (2021).


\bibitem{20}B. Allen, Phys. Rev. Lett., 63, 2017 (1989).

\bibitem{15}B. Allen, Gen. Relativ. Gravit., 22, 1447 (1990).



\bibitem{21}J. A. Frieman, D. D. Harari, \& G. C. Surpi, Phys. Rev. D, 50, 4895 (1994).





\bibitem{25}G. C. Surpi, D. D. Harari, \& J. A. Frieman, Astrophys. J., 464, 54 (1996).


\bibitem{26}A. S. Bolton, et al., Astrophys. J., 682, 964 (2008).

\bibitem{27}A. S. Bolton, et al., Astrophys. J., 684, 248 (2008).
\bibitem{28}D. P. Stark, et al., Mon. Not. R. Astron. Soc., 436, 1040 (2013).


\bibitem{29}C.-H. Lee, PASA, 34, e014 (2017).
\bibitem{30}C. Avestruz, et al., Astrophys. J, 877, 58 (2019).

\bibitem{31}S. AL Otaibi, PhD thesis, Machine learning methods for delay estimation in gravitationally lensed signals (2016).

\bibitem{32}A. G. de Bruyn, \& J.-P. Macquart, Astron. Astrophys., 574, A125 (2015).











\end{thebibliography}
\end{document}